\documentclass[10pt,aps,pra,twocolumn,showpacs,groupedaddress]{revtex4-1}
\usepackage[english]{babel}
\usepackage[T1]{fontenc}
\usepackage{natbib}
\usepackage{soul}
\usepackage{amssymb}
\usepackage{bm}
\usepackage{amsmath}
\usepackage[mathcal]{eucal}
\usepackage{color}
\usepackage{graphicx}
\usepackage{hyperref}
\usepackage{mathtools}
\usepackage{url}
\usepackage{tikz}
\usetikzlibrary{shapes}
\usetikzlibrary{decorations}
\usetikzlibrary{arrows}
\usetikzlibrary{matrix}

\newcommand {\nc} {\newcommand}
\nc {\beq} {\begin{eqnarray}}
\nc {\eeqn} [1] {\label{#1} \end{eqnarray}}
\nc {\eoln} [1] {\label{#1} \\}
\nc {\eol} {\nonumber \\}
\nc {\rref} [1] {(\ref{#1})}
\nc {\Eq} [1] {Eq.~(\ref{#1})}
\nc {\Ref} [1] {Ref.~\cite{#1}}
\nc {\la} {\mbox{$\langle$}}
\nc {\ra} {\mbox{$\rangle$}}
\nc {\dem} {\mbox{$\frac{1}{2}$}}
\nc {\cP} {\mathcal{P}}
\nc {\cN} {\mathcal{N}}
\nc {\ve} [1] {\mbox{\boldmath $#1$}}
\nc {\arrow} [2] {\mbox{$\mathop{\rightarrow}\limits_{#1 \rightarrow #2}$}}
\nc {\red}[1] {\textcolor{red}{#1}}
\nc {\mc}[3] {\multicolumn{#1}{#2}{#3}}
\nc {\dd}{\, \mathrm{d}}
\nc {\bs}[1]{\boldsymbol{#1}}
\nc {\ket}[1]{\vert #1 \rangle}
\nc {\bra}[1]{\langle #1 \vert}
\nc {\abs}[1]{\vert #1 \vert}
\nc {\avg}[1]{\langle #1 \rangle}
\nc {\braket}[2]{\langle #1 \vphantom{#2} \vert #2 \vphantom{#1} \rangle}
\nc {\abss}[1]{\left| #1 \right|}

\begin{document}

\title{Core correlation effects in multiconfiguration calculations of isotope shifts in Mg~I}
\author{Livio Filippin}
\email[]{Livio.Filippin@ulb.ac.be}
\affiliation{Chimie Quantique et Photophysique, Universit\'{e} libre de Bruxelles, B-1050 Brussels, Belgium}
\author{Michel Godefroid}
\email[]{mrgodef@ulb.ac.be}
\affiliation{Chimie Quantique et Photophysique, Universit\'{e} libre de Bruxelles, B-1050 Brussels, Belgium}
\author{J\"{o}rgen Ekman}
\email[]{jorgen.ekman@mah.se}
\affiliation{Group for Materials Science and Applied Mathematics, Malm\"{o} University, S-20506 Malm\"{o}, Sweden}
\author{Per J\"{o}nsson}
\email[]{per.jonsson@mah.se}
\affiliation{Group for Materials Science and Applied Mathematics, Malm\"{o} University, S-20506 Malm\"{o}, Sweden}

\date{\today}

\begin{abstract}
The present work reports results from systematic multiconfiguration Dirac-Hartree-Fock calculations of isotope shifts for several well-known transitions in neutral magnesium.
Relativistic normal and specific mass shift factors as well as the electronic probability density at the origin are calculated. Combining these electronic quantities with available nuclear data, energy and transition level shifts are determined for the $^{26}$Mg$-^{24}$Mg pair of isotopes. Different models for electron correlation are adopted. It is shown that although valence and core-valence models provide accurate values for the isotope shifts, the inclusion of core-core excitations in the computational strategy significantly improves the accuracy of the transition energies and normal mass shift factors.
\end{abstract}


\maketitle

%%%%%%%%%%%%%%%%%%%%%%%%%%%%%%%%%%%%%%%%%%%%%%%%%%%%%%%%%%%%%%%%%%%%%%%%%%%%%%%%%%%%%%%%%%%%%%%%%%%%%%%%%%%%%%%%%%%%%%%%%%
%%%%%%%%%%%%%%%%%%%%%%%%%%%%%%%%%%%%%%%%%%%%%%%%%%%%%%%%%%%%%%%%%%%%%%%%%%%%%%%%%%%%%%%%%%%%%%%%%%%%%%%%%%%%%%%%%%%%%%%%%%

\section{Introduction}
\label{sec:intro}

When the effects of the finite mass and the extended spatial charge distribution of the nucleus are taken into account in a Hamiltonian describing an atomic system, the isotopes of an element have different electronic energy levels~\cite{NGG13}. The isotope shift (IS) of spectral lines, which consists of the mass shift (MS) and the field shift (FS), plays a key role for extracting the  changes in the mean-square charge radius of the atomic nucleus~\cite{Ki84,CCF12,NLG15}.
For a given atomic transition $k$ with frequency~$\nu_k$, it is assumed that the electronic response of the atom to variations of the nuclear mass and charge distribution can be described by only two factors: the mass shift factor, $\Delta K_{k,MS}$, and the field shift factor, $F_k$, respectively. The observed IS, $\delta \nu_k^{A,A'}$, between any pair of isotopes with mass numbers $A$ and $A'$ is related to the difference in nuclear masses and in mean-square charge radii, $\delta \langle r^2 \rangle^{A,A'}$~\cite{Ki84,NGG13}.

We perform \textit{ab initio} calculations of IS electronic factors using the multiconfiguration Dirac-Hartree-Fock (MCDHF) method. This method is implemented in the \textsc{Ris}3 (relativistic isotope shift) module~\cite{NGG13}, designed for the revised version of the \textsc{Grasp}2K program package~\cite{JGB13}. The adopted computational scheme is based on the estimation of the expectation values of the one- and two-body recoil Hamiltonian for a given isotope, including relativistic corrections derived by Shabaev~\cite{Sh85,Sh88}, combined with the calculation of the total electron densities at the origin. Different correlation models are explored in a systematic way to determine a reliable computational strategy. This strategy is applied on neutral magnesium (Mg~I), which is one of the simplest and best-studied two-valence-electron atoms. As such, it is often used as a test ground for different methods of atomic calculations. In this paper we show that we can accurately calculate the isotope shift of some well known transitions in Mg~I, where experimental~\cite{SSM93,SHE06,HH78,NGT92,Ha79} and theoretical values~\cite{Ve87,JFG99,BFK05} are available for the $^{26}$Mg$-^{24}$Mg pair of isotopes.

In Sec.~\ref{sec:method}, the principles of the MCDHF method are summarised. In Sec.~\ref{sec:isotope_shifts}, the relativistic expressions of the MS and FS factors are recalled. Section~\ref{sec:AS} enumerates the studied transitions in Mg~I and presents the active space expansion strategy adopted for the electron correlation model. In Sec.~\ref{sec:results}, numerical results of the MS and FS factors are reported for each of the studied transitions, as well as transition energy shifts for the $^{26}$Mg$-^{24}$Mg pair of isotopes. Section~\ref{sec:conc} reports concluding remarks.

%%%%%%%%%%%%%%%%%%%%%%%%%%%%%%%%%%%%%%%%%%%%%%%%%%%%%%%%%%%%%%%%%%%%%%%%%%%%%%%%%%%%%%%%%%%%%%%%%%%%%%%%%%%%%%%%%%%%%%%%%%
%%%%%%%%%%%%%%%%%%%%%%%%%%%%%%%%%%%%%%%%%%%%%%%%%%%%%%%%%%%%%%%%%%%%%%%%%%%%%%%%%%%%%%%%%%%%%%%%%%%%%%%%%%%%%%%%%%%%%%%%%%

\section{Numerical method}
\label{sec:method}

The MCDHF method~\cite{Gr07}, as implemented in the \textsc{Grasp}2K program package~\cite{JGB13,JHF07}, is the fully relativistic counterpart of the non-relativistic multiconfiguration Hartree-Fock (MCHF) method~\cite{FTG07}. The MCDHF method is employed to obtain wave functions that are referred to as atomic state functions (ASF), i.e., approximate eigenfunctions of the Dirac-Coulomb Hamiltonian given by
\beq
\mathcal{H}_{\text{DC}} = \sum_{i=1}^N [c \bs{\alpha}_i \cdot \bs{p}_i + (\beta_i - 1)c^2 + V(r_i)] + \sum_{i<j}^N \frac{1}{r_{ij}},
\eeqn{eq_DC_Hamiltonian}
where $V(r_i)$ is the monopole part of the electron-nucleus interaction, $c$ is the speed of light and $\bs{\alpha}$ and $\beta$ are the $(4 \times 4)$ Dirac matrices. An ASF is given as an expansion over $jj$-coupled configuration state functions (CSFs), $\Phi(\gamma_{\nu}\Pi JM_J)$, with the same parity $\Pi$, total angular momentum  $J$ and $J_z$-projection $M_J$ quantum numbers:
\beq
\label{eq:wfn}
\vert \Psi(\gamma\,\Pi JM_J) \rangle = \sum_{\nu=1}^{N_{\text{CSFs}}} c_{\nu} \, \vert \Phi(\gamma_{\nu}\,\Pi JM_J) \rangle.
\eeqn{eq_ASF}

In the MCDHF method the radial functions, used to construct the CSFs, and the expansion coefficients $c_{\nu}$ are determined variationally so as to leave the energy functional
\beq
E = \sum_{\mu,\nu}^{N_{\text{CSFs}}} c_{\mu} c_{\nu} \langle \Phi(\gamma_{\mu}\,\Pi JM_J) \vert \mathcal{H}_{\text{DC}} \vert \Phi(\gamma_{\nu}\,\Pi JM_J) \rangle
\eeqn{eq_energy_functional}
stationary with respect to their variations. The resulting coupled radial equations are solved iteratively in the self-consistent field (SCF) procedure. Once radial functions have been determined, a configuration interaction (CI) calculation is performed over the set of configuration states, providing the expansion coefficients for building the potentials of the next iteration. The SCF and CI coupled processes are repeated until convergence of the total wave function~\rref{eq:wfn} is reached.

%%%%%%%%%%%%%%%%%%%%%%%%%%%%%%%%%%%%%%%%%%%%%%%%%%%%%%%%%%%%%%%%%%%%%%%%%%%%%%%%%%%%%%%%%%%%%%%%%%%%%%%%%%%%%%%%%%%%%%%%%%
%%%%%%%%%%%%%%%%%%%%%%%%%%%%%%%%%%%%%%%%%%%%%%%%%%%%%%%%%%%%%%%%%%%%%%%%%%%%%%%%%%%%%%%%%%%%%%%%%%%%%%%%%%%%%%%%%%%%%%%%%%

\section{Isotope shift theory}
\label{sec:isotope_shifts}

The main ideas of the IS theory are outlined here. More details can be found in the works by Shabaev \cite{Sh85,Sh88} and Palmer \cite{Pa88}, who pioneered the theory of the relativistic mass shift used in the present work. Gaidamauskas \textit{et al.}~\cite{GNR11} derived the tensorial form of the relativistic recoil operator, implemented in \textsc{Ris}3~\cite{NGG13}.

\subsection{Mass shift}
\label{subsec:mass_shift}

The finite mass of the nucleus gives rise to a recoil effect, called the mass shift (MS). The nuclear recoil corrections within the $(\alpha Z)^4m^2/M$ approximation~\cite{Sh85,Sh88} are obtained by evaluating the expectation values of the one- and two-body recoil Hamiltonian for a given isotope,
\beq
\mathcal{H}_{\text{MS}} = \frac{1}{2M} \sum_{i,j}^{N} \left( \bs{p}_i \cdot \bs{p}_j - \frac{\alpha Z}{r_i} (\bs{\alpha}_i + \frac{(\bs{\alpha}_i \cdot \bs{r}_i) \bs{r}_i}{r_i^2}) \cdot \bs{p}_j \right), \eol
\eeqn{eq_H_MS}
where $M$ stands for the mass of the nucleus. Separating the one-body $(i=j)$ and two-body $(i\neq j)$ terms that, respectively, constitute the normal mass shift (NMS) and specific mass shift (SMS) contributions, the Hamiltonian \rref{eq_H_MS} can be written
\beq
\mathcal{H}_{\text{MS}}=\mathcal{H}_{\text{NMS}}+\mathcal{H}_{\text{SMS}}.
\eeqn{eq_H_NMS_SMS} 

The NMS and SMS mass-independent $K$ factors are defined by the following expressions:
\beq
K_{\text{NMS}} \equiv M \langle \Psi \vert \mathcal{H}_{\text{NMS}} \vert \Psi \rangle,
\eeqn{eq_K_NMS}
and
\beq
K_{\text{SMS}} \equiv M \langle \Psi \vert \mathcal{H}_{\text{SMS}} \vert \Psi \rangle.
\eeqn{eq_K_SMS}

For a transition IS, one needs to consider the variation of the mass shift factor from one level to another. The corresponding line frequency isotope MS between two isotopes, $A$ and $A'$, is written as the sum of the NMS and SMS contributions,
\beq
\delta \nu_{k,\text{MS}}^{A,A'} & \equiv & \nu_{k,\text{MS}}^{A} - \nu_{k,\text{MS}}^{A'} \eol
& = & \delta \nu_{k,\text{NMS}}^{A,A'} + \delta \nu_{k,\text{SMS}}^{A,A'},
\eeqn{eq_delta_nu_MS}
with
\beq
\delta \nu_{k,\text{MS}}^{A,A'} & = & \left( \frac{1}{M} - \frac{1}{M'} \right) \frac{\Delta K_{k,\text{MS}}}{h} \eol
& = & \left( \frac{1}{M} - \frac{1}{M'} \right) \Delta \tilde{K}_{k,\text{MS}}.
\eeqn{eq_delta_nu_MS_2}
Here $\Delta K_{k,\text{MS}}=(K_{\text{MS}}^u-K_{\text{MS}}^l)$ is the difference of the $K_{\text{MS}}=K_{\text{NMS}}+K_{\text{SMS}}$ factors of the upper ($u$) and lower ($l$) levels involved in the transition $k$. For the $\Delta \tilde{K}$ factors the unit (GHz~u) is often used in the literature. As far as the conversion factors are concerned, we use $\Delta K_{k,\text{MS}}\,[m_eE_{\text{h}}]=3609.4824\,\Delta \tilde{K}_{k,\text{MS}}\,[\text{GHz~u}]$.

%%%%%%%%%%%%%%%%%%%%%%%%%%%%%%%%%%%%%%%%%%%%%%%%%%%%%%%%%%%%%%%%%%%%%%%%%%%%%%%%%%%%%%%%%%%%%%%%%%%%%%%%%%%%%%%%%%%%%%%%%%

\subsection{Field shift}
\label{subsec:field_shift}

Neglecting terms of higher order than $\delta \langle r^2 \rangle$ in the Seltzer moment~\cite{Se69}
\beq
\lambda^{A,A'} = \delta \langle r^2 \rangle^{A,A'} + b_1 \delta \langle r^4 \rangle^{A,A'} + b_2 \delta \langle r^6 \rangle^{A,A'} + \cdots \eol
\eeqn{eq_Seltzer_moment}
the line frequency shift in the transition $k$ arising from the difference in nuclear charge distributions between two isotopes, $A$ and $A'$, can be written as~\cite{FBH95,TFR85,BBP87}
\beq
\delta \nu_{k,\text{FS}}^{A,A'} & \equiv & \nu_{k,\text{FS}}^{A} - \nu_{k,\text{FS}}^{A'} \eol
& = & F_k \, \delta \langle r^2 \rangle^{A,A'}.
\eeqn{eq_delta_nu_FS}
In the expression above $\delta \langle r^2 \rangle^{A,A'} \equiv \langle r^2 \rangle^{A}-\langle r^2 \rangle^{A'}$, and $F_{k}$ is the line electronic factor given by
\beq
F_k = \frac{2\pi}{3h} Z \left( \frac{e^2}{4\pi \epsilon_0} \right) \Delta \vert \Psi(0) \vert_k^2,
\eeqn{eq_F_k}
which is proportional to the change of the total electronic probability density at the origin between level $l$ and level~$u$,
\beq
\Delta \vert \Psi(0) \vert_k^2 & = & \Delta \rho_k^e(\bs{0}) \eol
& = & \rho_u^e(\bs{0}) - \rho_l^e(\bs{0}).
\eeqn{eq_Delta_Psi_0}

%%%%%%%%%%%%%%%%%%%%%%%%%%%%%%%%%%%%%%%%%%%%%%%%%%%%%%%%%%%%%%%%%%%%%%%%%%%%%%%%%%%%%%%%%%%%%%%%%%%%%%%%%%%%%%%%%%%%%%%%%%

\subsection{Total isotope shift}
\label{subsec:total_shift}

The total line frequency shift is obtained by merely adding the MS, \rref{eq_delta_nu_MS}, and FS, \rref{eq_delta_nu_FS}, contributions:
\beq
\delta \nu_k^{A,A'} & = & \overbrace{\delta \nu_{k,\text{NMS}}^{A,A'} + \delta \nu_{k,\text{SMS}}^{A,A'}}^{\delta \nu_{k,\text{MS}}^{A,A'}} + \delta \nu_{k,\text{FS}}^{A,A'} \eol
& = & \left( \frac{1}{M} - \frac{1}{M'} \right) \Delta \tilde{K}_{k,\text{MS}} + F_k \, \delta \langle r^2 \rangle^{A,A'}.
\eeqn{eq_IS}

\clearpage

%%%%%%%%%%%%%%%%%%%%%%%%%%%%%%%%%%%%%%%%%%%%%%%%%%%%%%%%%%%%%%%%%%%%%%%%%%%%%%%%%%%%%%%%%%%%%%%%%%%%%%%%%%%%%%%%%%%%%%%%%%
%%%%%%%%%%%%%%%%%%%%%%%%%%%%%%%%%%%%%%%%%%%%%%%%%%%%%%%%%%%%%%%%%%%%%%%%%%%%%%%%%%%%%%%%%%%%%%%%%%%%%%%%%%%%%%%%%%%%%%%%%%

\onecolumngrid

\begin{table}[ht!]
\caption{\small{Reference configurations for the lower and upper states of the studied transitions in Mg~I. The MR-cutoff values, $\varepsilon_{\text{MR}}$, determine the set of CSFs in the MR space. $N_{\text{CSFs}}$ represents the number of CSFs describing each MR space.}}
\begin{center}
\begin{tabular}{l c l l r}
\hline
\hline
Transition                                        & ~~~$\varepsilon_{\text{MR}}$~~~ & $J^\Pi$ & Reference configurations                                                             & $N_{\text{CSFs}}$ \\
\hline
$3s^{2}~^{1}S_{0} \rightarrow 3s3p~^{3}P^{o}_{1}$ & 0.01                            & $0^+$   & [Ne]$\{3s^2,3s4s,3p^2,3p4p,3d^2,4s^2,4p^2\}$                                         & 11                \\
\vspace{0.2cm}
                                                  &                                 & $1^-$   & [Ne]$\{3s3p,3s4p,3p3d,3p4s,3d4p,4s4p\}$                                              & 14                \\
$3s^{2}~^{1}S_{0} \rightarrow 3s3p~^{1}P^{o}_{1}$ & 0.01                            & $0^+$   & [Ne]$\{3s^2,3s4s,3p^2,3p4p,3d^2,3d4d,4s^2,4p^2\}$                                    & 12                \\
\vspace{0.2cm}
                                                  &                                 & $1^-$   & [Ne]$\{3s3p,3s4p,3p3d,3p4s,3p4d,3d4p,3d4f,4s4p\}$                                    & 18                \\
$3s3p~^{3}P^{o}_{1} \rightarrow 3s4s~^{3}S_{1}$   & 0.005                           & $1^-$   & [Ne]$\{3s3p,3s4p,3p3d,3d4p,3d4f,4s4p,3p5s,4p5s\}$                                    & 18                \\
\vspace{0.2cm}
                                                  &                                 & $1^+$   & [Ne]$\{3s4s,3p4p,3d4d,3s5s,4s5s\}$                                                   & 10                \\
$3s3p~^{3}P^{o}_{1} \rightarrow 3p^2~^{3}P_{0}$   & 0.01                            & $1^-$   & [Ne]$\{3s3p,3s4p,3p3d,3d4p,4s4p\}$                                                   & 10                \\
\vspace{0.2cm}
                                                  &                                 & $0^+$   & [Ne]$\{3s^2,3s4s,3s5s,3s6s,3p^2,3p4p,3d^2,4s^2,4s5s,4s6s,4p^2,4f^2,5s6s,5s^2,6s^2\}$ & 19                \\      
$3s3p~^{3}P^{o}_{1} \rightarrow 3s3d~^{3}D_{1}$   & 0.01                            & $1^-$   & [Ne]$\{3s3p,3p3d,3p4s,3p4d,3d4p,4s4p,4p4d\}$                                         & 15                \\
\vspace{0.2cm}
                                                  &                                 & $1^+$   & [Ne]$\{3s3d,3p4p,3p4f,3d4s,4s4d,4p4f\}$                                              & 9                 \\
$3s3p~^{3}P^{o}_{1} \rightarrow 3s4d~^{3}D_{1}$   & 0.01                            & $1^-$   & [Ne]$\{3s3p,3p3d,3p4d,3d4p,3d4f,4s4p,4p4d,3p5s,3p6s,4p6s\}$                          & 19                \\
                                                  &                                 & $1^+$   & [Ne]$\{3s3d,3s4s,3s4d,3s5s,3p4p,3p4f,3d4s,3d4d,3p^2,3d^2,4s4d,4s5s,4p4f,4p^2,$       & 23                \\
\vspace{0.2cm}
                                                  &                                 &         & \quad \:\:\:\: $4f^2,3s6s,3d6s,4s6s,4d6s,5s6s\}$                                     &                   \\
$3s3p~^{1}P^{o}_{1} \rightarrow 3s4d~^{1}D_{2}$   & 0.025                           & $1^-$   & [Ne]$\{3s3p,3s4p,3p3d,3p4s,3p4d,3d4p\}$                                              & 12                \\
                                                  &                                 & $2^+$   & [Ne]$\{3s3d,3s4d,3p^2,3p4p,3p4f\}$                                                   & 11                \\
\hline
\hline
\end{tabular}
\end{center}
\label{table_MR_composition}
\end{table}

\twocolumngrid

\section{Active space expansion}
\label{sec:AS}

\begin{figure}[ht!]
\begin{center}
\begin{tikzpicture}[scale=0.5,level/.style={ultra thick},trans/.style={thick,->,shorten >=2.5pt,shorten <=2.5pt,>=stealth}]
\node at (-3,-1) {$J=0$};
\node at (3.25,-1) {$J=1$};
\node at (9.5,-1) {$J=2$};

\draw[level] (-5,0) -- (-1,0) node[midway,above] {$3s^{2}~^{1}S_{0}$};
\draw[level] (-5,11.5626) -- (-1,11.5626) node[midway,above] {$3p^{2}~^{3}P_{0}$};
\draw[level] (0,4.3740) -- (6.5,4.3740) node[midway,above] {$3s3p~^{3}P^{o}_{1}$};
\draw[level] (0,7.0102) -- (6.5,7.0102) node[midway,above] {$3s3p~^{1}P^{o}_{1}$};
\draw[level] (0,8.2394) -- (6.5,8.2394) node[midway,above] {$3s4s~^{3}S_{1}$};
\draw[level] (0,9.5914) -- (6.5,9.5914) node[midway,above] {$3s3d~^{3}D_{1}$};
\draw[level] (0,10.8384) -- (6.5,10.8384) node[midway,above] {$3s4d~^{3}D_{1}$};
\draw[level] (7.5,10.6270) -- (11.5,10.6270) node[midway,above] {$3s4d~^{1}D_{2}$};

\draw[trans] (-1.75,0) -- (3.25,7.0102) node[midway,sloped,below] {\scriptsize{285.3 nm}};
\draw[trans] (-1,0) -- (3.25,4.3740) node[midway,sloped,below] {\scriptsize{457.2 nm}};
\draw[trans] (0.75,4.3740) -- (0.75,8.2394) node[midway,xshift=-0.9em,sloped,below] {\scriptsize{517.4 nm}};
\draw[trans] (0,4.3740) -- (-3,11.5626) node[midway,sloped,below] {\scriptsize{278.2 nm}};
\draw[trans] (5,4.3740) -- (5,9.5914) node[midway,xshift=-1.9em,sloped,below] {\scriptsize{383.3 nm}};
\draw[trans] (5.75,4.3740) -- (5.75,10.8384) node[midway,xshift=-2.85em,sloped,below] {\scriptsize{309.4 nm}};
\draw[trans] (6.5,7.0102) -- (9.5,10.6270) node[midway,sloped,below] {\scriptsize{553.0 nm}};
\end{tikzpicture}
\end{center}
\caption{\small{Schematic diagram of the Mg~I transitions.}}
\label{transitions_diagram}
\end{figure}
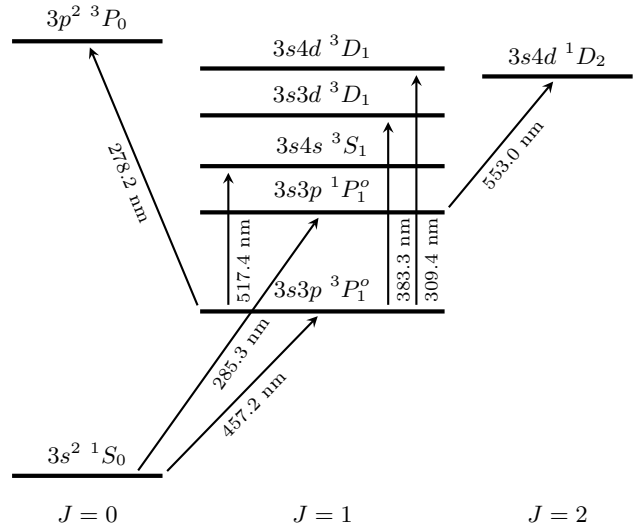

The transitions in Mg~I considered in the present work are the following (see \figurename{~\ref{transitions_diagram}}): $3s^{2}~^{1}S_{0} \rightarrow 3s3p~^{3}P^{o}_{1}$ (457.2~nm), $3s^{2}~^{1}S_{0} \rightarrow 3s3p~^{1}P^{o}_{1}$ (285.3~nm), $3s3p~^{3}P^{o}_{1} \rightarrow 3s4s~^{3}S_{1}$ (517.4~nm), $3s3p~^{3}P^{o}_{1} \rightarrow 3p^2~^{3}P_{0}$ (278.2~nm), $3s3p~^{3}P^{o}_{1} \rightarrow 3s3d~^{3}D_{1}$ (383.3~nm), $3s3p~^{3}P^{o}_{1} \rightarrow 3s4d~^{3}D_{1}$ (309.4~nm) and $3s3p~^{1}P^{o}_{1} \rightarrow 3s4d~^{1}D_{2}$ (553.0~nm).

To effectively capture electron correlation, CSFs of a particular symmetry ($J$) and parity ($\Pi$) are generated through excitations within an active set of orbitals occupied in the reference configurations and non-occupied virtual orbitals. From hardware and software limitations, it is impossible to use complete active space (CAS) wave functions that would include all CSFs with the appropriate $J$ and $\Pi$ for a given orbital active set. Hence the CSF expansions have to be constrained so that major correlation excitations are taken into account~\cite{NLG15}.

Single (S) and double (D) substitutions are performed on a multireference (MR) set, which contains the CSFs that have large expansion coefficients and account for the major correlation effects. These SD-MR substitutions take into account valence-valence (VV), core-valence (CV) as well as core-core (CC) correlations. The VV correlation model only allows SD substitutions from valence orbitals, while the VV+CV correlation model considers SrD substitutions (single and restricted double) from core and valence orbitals, limiting the excitations to a maximum of one hole in the core. By contrast, the VV+CV+CC correlation model allows all SD substitutions from core and valence orbitals.

Within this approach, a common orbital basis set is chosen for the lower and upper states of each transition. The reference states are obtained using a valence-CAS procedure: SD substitutions are performed within the $n=3,4$ valence orbitals, also including the $5s$ or both $5s$ and $6s$ orbitals in the active space for some transitions (see \tablename{~\ref{table_MR_composition}}). The $5s$ and $6s$ orbitals are added to account for states belonging to lower configurations with the same $J$ and $\Pi$ in the optimisation of the energy functional.

An SCF procedure is then applied to the resulting CSFs, providing the orbital set and the expansion coefficients. Due to limited computer resources, such a valence-CAS multireference set would be too large for subsequent calculations when the active orbital set increases. Hence, for reducing the size of the MR set, only the CSFs whose squared expansion coefficients are larger than a given MR-cutoff are kept, i.e., $c_\nu^2>\varepsilon_{\text{MR}}$. For each transition, the $\varepsilon_{\text{MR}}$ values and the resulting MR sets are listed in \tablename{~\ref{table_MR_composition}}, for the lower and upper states.

The $1s$ orbital is kept closed in all subsequent calculations, i.e., no substitution from this orbital is allowed. Tests show that opening the $1s$ orbital does not affect the MS and FS factors to any notable extent. Only orbitals occupied in the single configuration DHF approximation are treated as spectroscopic, and the occupied reference orbitals are frozen in all subsequent calculations. The $J$-levels belonging to a given term are optimised simultaneously with standard weights through the Extended Optimal Level (EOL) scheme~\cite{DGJ89} and the set of virtual orbitals is increased layer by layer.

For a given transition, the optimisation procedure is summarised as follows:
\begin{enumerate}
\item Perform simultaneous calculations for the lower and upper states of the transition, using a MR set consisting of CSFs with the form $2s^22p^6nln'l'~^{2S+1}L_J$ with $n,n'=3,4$ (+$5s$ or $5s,6s$) and $l,l'=s,p,d,f$. Optimise all orbitals simultaneously. These CSFs account for a fair amount of the VV correlation.
\item Keep the orbitals fixed from step 1, and optimise an orbital basis layer by layer up to $n=8h$ for both states of the transition, described by CSFs with respective $J^\Pi$ symmetries. These CSFs are obtained by SD-MR substitutions with the restriction that there is at most one excitation from the $2s^22p^6$ core.
\item Perform a CI calculation on the CSFs expansion with the $J^\Pi$ symmetry of both states, describing VV, CV and CC correlation obtained by SD-MR substitutions to the orbital basis from step 2.
\item Keep the orbitals up to $8h$ fixed from step 2 and optimise one additional layer of orbitals using CC substitutions from the Mg$^{2+}$ $2s^22p^6$ ($Z=12$) core.
The orbitals of this additional layer target CC correlation, and are therefore contracted.
\item Perform a CI calculation on the CSFs expansion with the $J^\Pi$ symmetry of both states of the transition, describing VV, CV and CC correlation obtained by SD-MR substitutions to the orbital basis from step 4 ($n=8h~+$ the additional layer).
\end{enumerate}
Following the procedure in steps 1-2 or 1-5 respectively yield results labelled `CV' or `CC' in Tables~\ref{table_E_KNMS_KSMS_F} and \ref{table_IS_NMS_SMS_FS}.

The CC effects are more balanced with a common orbital basis for describing both upper and lower states, resulting in more accurate transition energies, as mentioned in \Ref{Ve87}.

The CSFs expansions become significantly large when CC correlations are taken into account, counting up to $2,000,000$~CSFs. Hence, applying an SCF procedure to such amount of CSFs takes too much computing time. This justifies the use of the CI method at that stage.

The effect of adding the Breit interaction to the Dirac-Coulomb Hamiltonian, \rref{eq_DC_Hamiltonian}, is found to be much smaller than the uncertainty in the transition IS factors with respect to the correlation model. This interaction has therefore been neglected in the procedure.

%%%%%%%%%%%%%%%%%%%%%%%%%%%%%%%%%%%%%%%%%%%%%%%%%%%%%%%%%%%%%%%%%%%%%%%%%%%%%%%%%%%%%%%%%%%%%%%%%%%%%%%%%%%%%%%%%%%%%%%%%%
%%%%%%%%%%%%%%%%%%%%%%%%%%%%%%%%%%%%%%%%%%%%%%%%%%%%%%%%%%%%%%%%%%%%%%%%%%%%%%%%%%%%%%%%%%%%%%%%%%%%%%%%%%%%%%%%%%%%%%%%%%

\section{Numerical results}
\label{sec:results}

In this section, MS and FS electronic factors, $\Delta \tilde{K}_{k,\text{MS}}$ and $F_k$, as well as total IS, $\delta \nu_k^{26,24}$, given by
\beq
\delta \nu_k^{26,24} = \left( \frac{1}{M_{26}} - \frac{1}{M_{24}} \right) \Delta \tilde{K}_{k,\text{MS}} + F_k \, \delta \langle r^2 \rangle^{26,24} \eol
\eeqn{eq_Delta_nu_26_24}
of the $^{26}$Mg$-^{24}$Mg pair of isotopes are calculated for the studied transitions in Mg~I.

Nuclear masses ($M$) are calculated by substracting the mass of the electrons and the binding energy from the atomic mass ($M_{\text{atom}}$), using the formula
\beq
M(A,Z) = M_{\text{atom}}(A,Z) - Zm_e + B_{\text{el}}(Z),
\eeqn{eq_nuclear_mass}
where the total electronic binding energy (in eV) is estimated using~\cite{HAC76,LPT03}
\beq
B_{\text{el}}(Z) = 14.4381 Z^{2.39} + 1.55468 \times 10^{-6} Z^{5.35}.
\eeqn{eq_binding_energy}
Atomic masses are provided in~\cite{CST12}. The resulting values of the nuclear masses are respectively
\beq
M_{26} = 25.97601589 \, \text{u}
\eeqn{eq_M_26}
and
\beq
M_{24} = 23.97846462 \, \text{u}.
\eeqn{eq_M_24}

The NMS factor, $\Delta \tilde{K}_{k,\text{NMS}}$, can be approximated through the scaling law
\beq
\Delta \tilde{K}_{k,\text{NMS}} \approx -m_e \nu_k^{\text{exp}},
\eeqn{eq_sl1}
where $m_e$ is the mass of the electron and $\nu_k^{\text{exp}}$ is the experimental transition energy of transition $k$, available in the NIST database~\cite{KRR15}. The transition NMS is then deduced from \Eq{eq_sl1} using expressions \rref{eq_delta_nu_MS} and \rref{eq_delta_nu_MS_2}, i.e.,
\beq
\delta \nu_{k,\text{NMS}}^{26,24} \approx \left( \frac{m_e}{M_{24}} - \frac{m_e}{M_{26}} \right) \nu_k^{\text{exp}}.
\eeqn{eq_sl2}

The reliability of the FS values obtained with the \textit{ab initio} electronic $F_k$ factor, \rref{eq_F_k}, is a function of the accuracy of the calculations, but also of the level of confidence on the nuclear data $\delta \langle r^2 \rangle^{A,A'}$. Values compiled by Angeli and Marinova~\cite{AM13} provide the mean-square charge radii difference between $^{26}$Mg and $^{24}$Mg:
\beq
\delta \langle r^2 \rangle^{26,24} & \equiv & \langle r^2 \rangle^{26} - \langle r^2 \rangle^{24} \eol
& = & -0.1419 \, \text{fm}^2.
\eeqn{eq_delta_r2_26_24}

\clearpage

\onecolumngrid

\begin{table}[ht!]
\caption{\small{Level MS factors, $K_{\text{NMS}}$ and $K_{\text{SMS}}$ (in $m_eE_{\text{h}}$), and the electronic probability density at the origin, $\rho^e(\bs{0})$ (in $a_0^{-3}$), as a function of the increasing active space for the $3s^{2}~^{1}S_{0} \rightarrow 3s3p~^{1}P^{o}_{1}$ transition in Mg~I. $\Delta^u_l$ stands for the difference between the values of the upper level and the lower level. Results are obtained with a MR-cutoff $\varepsilon_{\text{MR}}=0.01$.}}
\begin{center}
\begin{tabular}{l c c c c c c c c c c c c c c c c}
\hline
\hline
               &          & & & \mc{3}{c}{$K_{\text{NMS}}$ ($m_eE_{\text{h}}$)} & & & \mc{3}{c}{$K_{\text{SMS}}$ ($m_eE_{\text{h}}$)} & & & \mc{3}{c}{$\rho^e(\bs{0})$ ($a_0^{-3}$)} \\
\cline{5-7} \cline{10-12} \cline{15-17}
Active space   & Notation & & & lower      & upper      & $\Delta^u_l$          & & & lower      & upper      & $\Delta^u_l$          & & & lower       & upper       & $\Delta^u_l$ \\
\hline
VV model       &          & & &            &            &                       & & &            &            &                       & & &             &             &              \\
\vspace{0.2cm}
$4s4p4d4f$     & VV $4f$  & & & $199.6791$ & $199.4764$ & $-0.2027$             & & & $-27.5167$ & $-27.4115$ & $0.1052$              & & & $1157.2594$ & $1156.4404$ & $-0.8190$    \\
VV+CV model    &          & & &            &            &                       & & &            &            &                       & & &             &             &              \\
$5s5p5d5f5g$   & CV $5g$  & & & $199.6023$ & $199.4798$ & $-0.1225$             & & & $-27.3690$ & $-27.3370$ & $0.0320$              & & & $1157.5654$ & $1156.5885$ & $-0.9769$    \\
$6s6p6d6f6g6h$ & CV $6h$  & & & $199.6306$ & $199.4951$ & $-0.1355$             & & & $-27.3645$ & $-27.3266$ & $0.0379$              & & & $1157.6171$ & $1156.6106$ & $-1.0065$    \\
$7s7p7d7f7g7h$ & CV $7h$  & & & $199.6337$ & $199.4974$ & $-0.1363$             & & & $-27.3600$ & $-27.3288$ & $0.0312$              & & & $1157.6394$ & $1156.6151$ & $-1.0243$    \\
\vspace{0.2cm}
$8s8p8d8f8g8h$ & CV $8h$  & & & $199.6338$ & $199.4974$ & $-0.1364$             & & & $-27.3518$ & $-27.3239$ & $0.0279$              & & & $1157.6481$ & $1156.6246$ & $-1.0235$    \\
VV+CV+CC model &          & & &            &            &                       & & &            &            &                       & & &             &             &              \\
$8s8p8d8f8g8h$ & CC $8h$  & & & $199.9113$ & $199.7546$ & $-0.1567$             & & & $-24.3327$ & $-24.2797$ & $0.0530$              & & & $1157.6349$ & $1156.6514$ & $-0.9835$    \\
$9s9p9d9f9g9h$ & CC $9h$  & & & $199.9401$ & $199.7829$ & $-0.1572$             & & & $-24.3200$ & $-24.2668$ & $0.0532$              & & & $1157.6521$ & $1156.6665$ & $-0.9856$    \\
\hline
\hline
\end{tabular}
\end{center}
\label{table_KNMS_KSMS_rho}
\end{table}

\twocolumngrid

Let us first study the convergence of the level MS factors, $K_{\text{NMS}}$ and $K_{\text{SMS}}$ (in $m_eE_{\text{h}}$), and the electronic probability density at the origin, $\rho^e(\bs{0})$ (in $a_0^{-3}$), of a given transition as a function of the increasing active space. \tablename{~\ref{table_KNMS_KSMS_rho}} displays the values for the $3s^{2}~^{1}S_{0} \rightarrow 3s3p~^{1}P^{o}_{1}$ transition, with a MR-cutoff $\varepsilon_{\text{MR}}$ equal to 0.01. Within each correlation model, the active space is extended until convergence of the differential results $\Delta^u_l$ is obtained.

For $K_{\text{NMS}}$, adding the $n=5$ layer of orbitals optimised on VV and CV correlations (denoted as `CV~$5g$' in \tablename{~\ref{table_KNMS_KSMS_rho}}) slightly shifts the value for the lower level, $3s^{2}~^{1}S_{0}$, while the value for the upper level, $3s3p~^{1}P^{o}_{1}$, remains nearly constant. However, this small variation leads to a significant modification ($40\%$) of the differential value, $\Delta K_{\text{NMS}}$. The convergence of the results is achieved by adding the successive layers within the VV+CV model, when the active space includes the $n=8$ correlation layer (denoted as `CV~$8h$'). Adding CC correlations through the CI computation described in step 3 of Sec.~\ref{sec:AS} (denoted as `CC~$8h$') shifts both level values, and hence does not drastically modify $\Delta K_{\text{NMS}}$ ($15\%$). The convergence is obtained for $\Delta K_{\text{NMS}}$ within the VV+CV+CC model, with the procedure of steps 4 and~5 (denoted as `CC~$9h$').

The situation is different for $K_{\text{SMS}}$. Adding the $n=5$ layer in the active space (`CV~$5g$') modifies both level and differential values. The convergence for $\Delta K_{\text{SMS}}$ within the VV+CV model is slower than for $\Delta K_{\text{NMS}}$. It is only obtained when the $n=9$ correlation layer is included, where $\Delta K_{\text{SMS}}= 0.0275 \, m_eE_{\text{h}}$. It is indeed well known that the SMS factor is more sensitive to correlation effects than the NMS factor, as expected from the two-body nature of the SMS operator. However, the inclusion of this last VV+CV correlation layer does not affect the results when CC correlations are added, and hence is not considered in this work. The procedure of step 3 (`CC~$8h$') leads to a drastic change in the level values, and also in the $\Delta K_{\text{SMS}}$ value~($90\%$). Within the VV+CV+CC model, the procedure of steps 4 and 5 (`CC~$9h$') leads to the convergence of $\Delta K_{\text{SMS}}$.

The convergence is smoother for $\rho^e(\bs{0})$ compared with $K_{\text{SMS}}$, as expected from the one-body nature of the density operator, like the NMS operator. The $\Delta \rho^e(\bs{0})$ value converges within the VV+CV model (`CV~$8h$'). Adding CC correlations in step 3 (`CC~$8h$') does not significantly affect both level and differential values. Within the VV+CV+CC model, the procedure of steps 4 and 5 (`CC~$9h$') leads to the convergence of $\Delta \rho^e(\bs{0})$.

A look at both the MS and FS factors displayed in \tablename{~\ref{table_KNMS_KSMS_rho}} shows that small variations in the level values due to correlation effects can lead to a significant variation in the differential values, $\Delta^u_l$. This illustrates how sensitive these electronic factors are, and hence how challenging it is to obtain reliable values with such a computational approach. This observation is general for all other transitions studied in this work.

\begin{figure}[h!]
\begin{minipage}[c]{.49\linewidth}
\begin{center}
\includegraphics[scale=0.4]{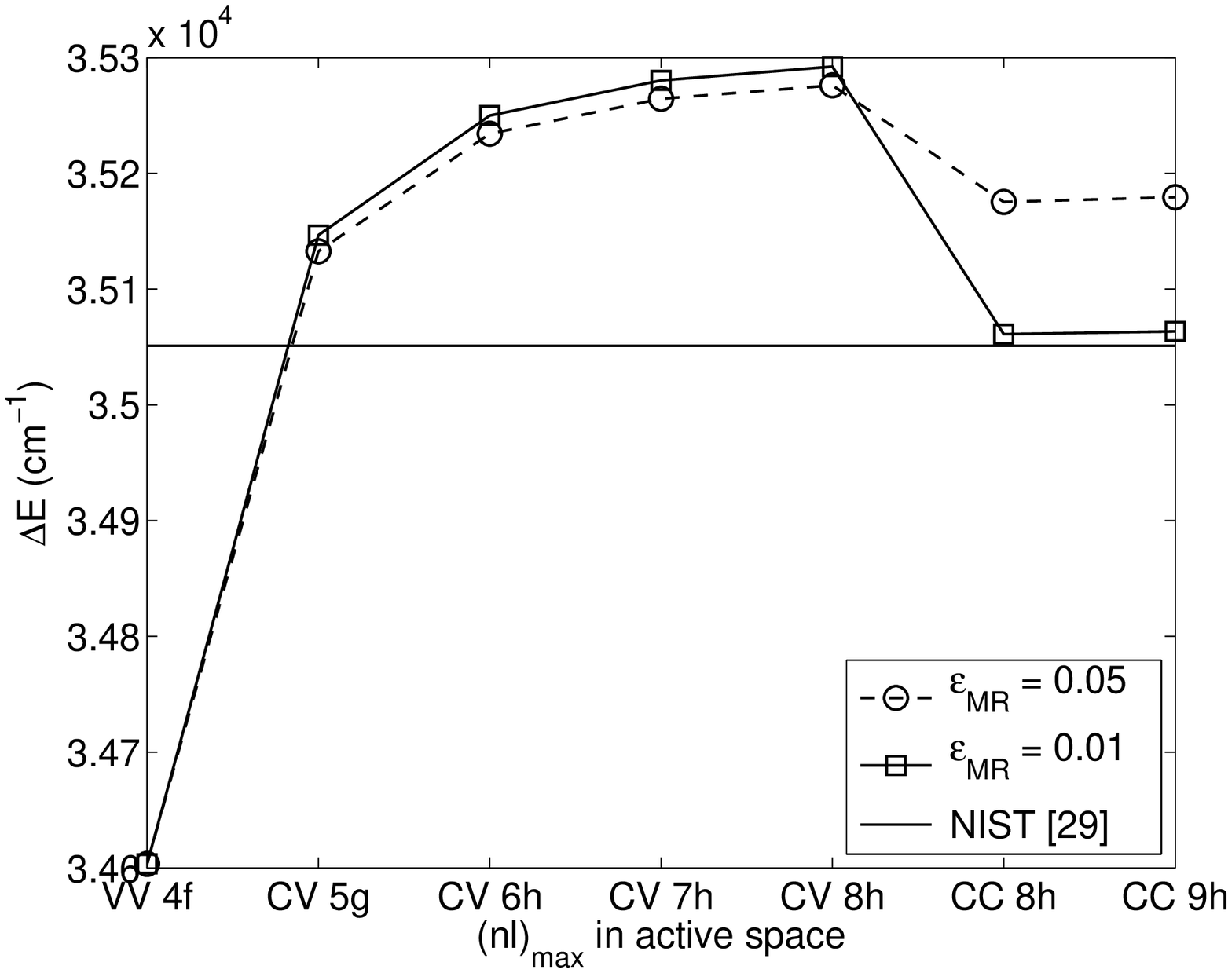}
\end{center}
\end{minipage}
\begin{minipage}[c]{.49\linewidth}
\begin{center}
\includegraphics[scale=0.4]{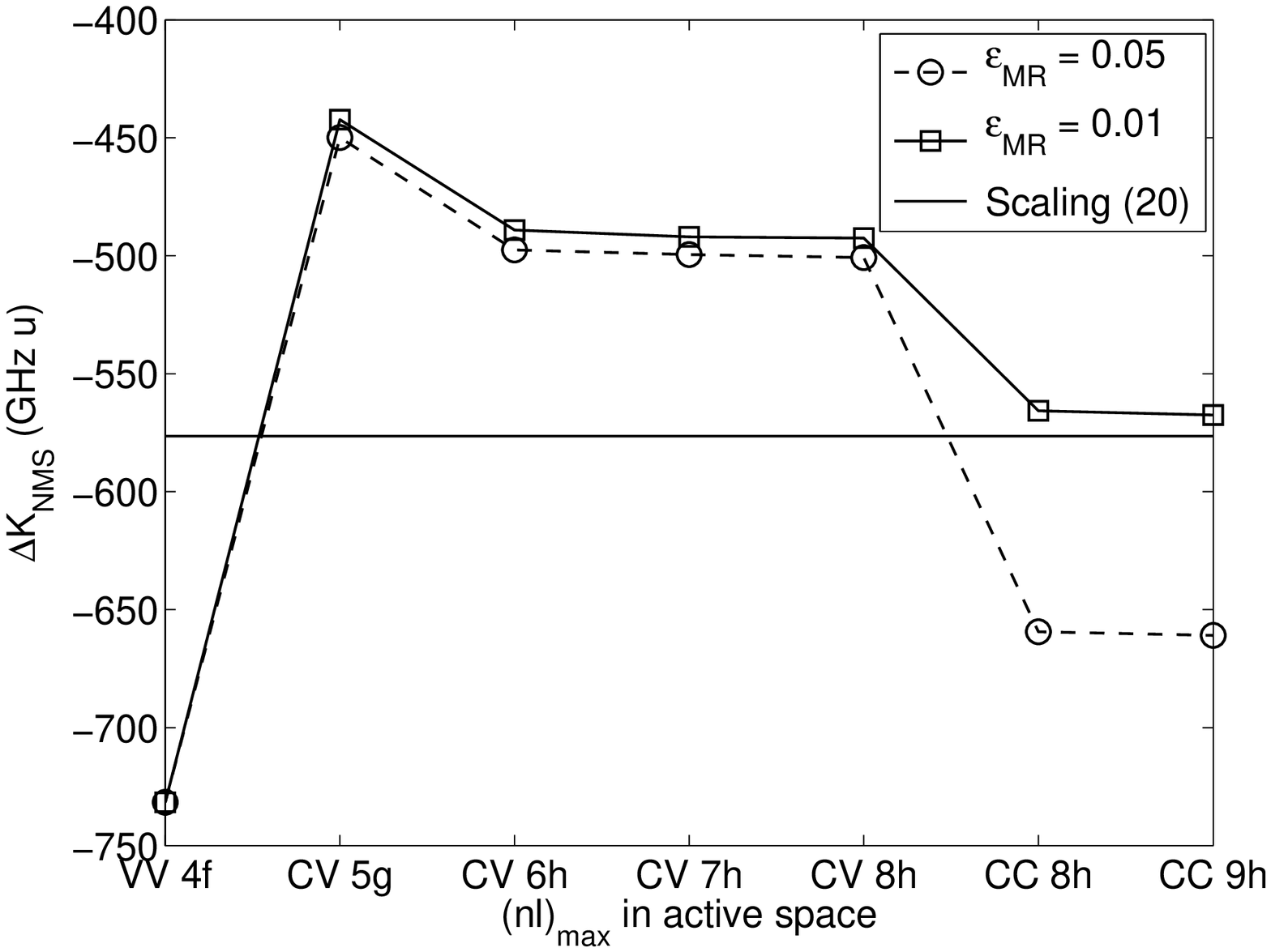}
\end{center}
\end{minipage}

\begin{minipage}[c]{.49\linewidth}
\begin{center}
\includegraphics[scale=0.4]{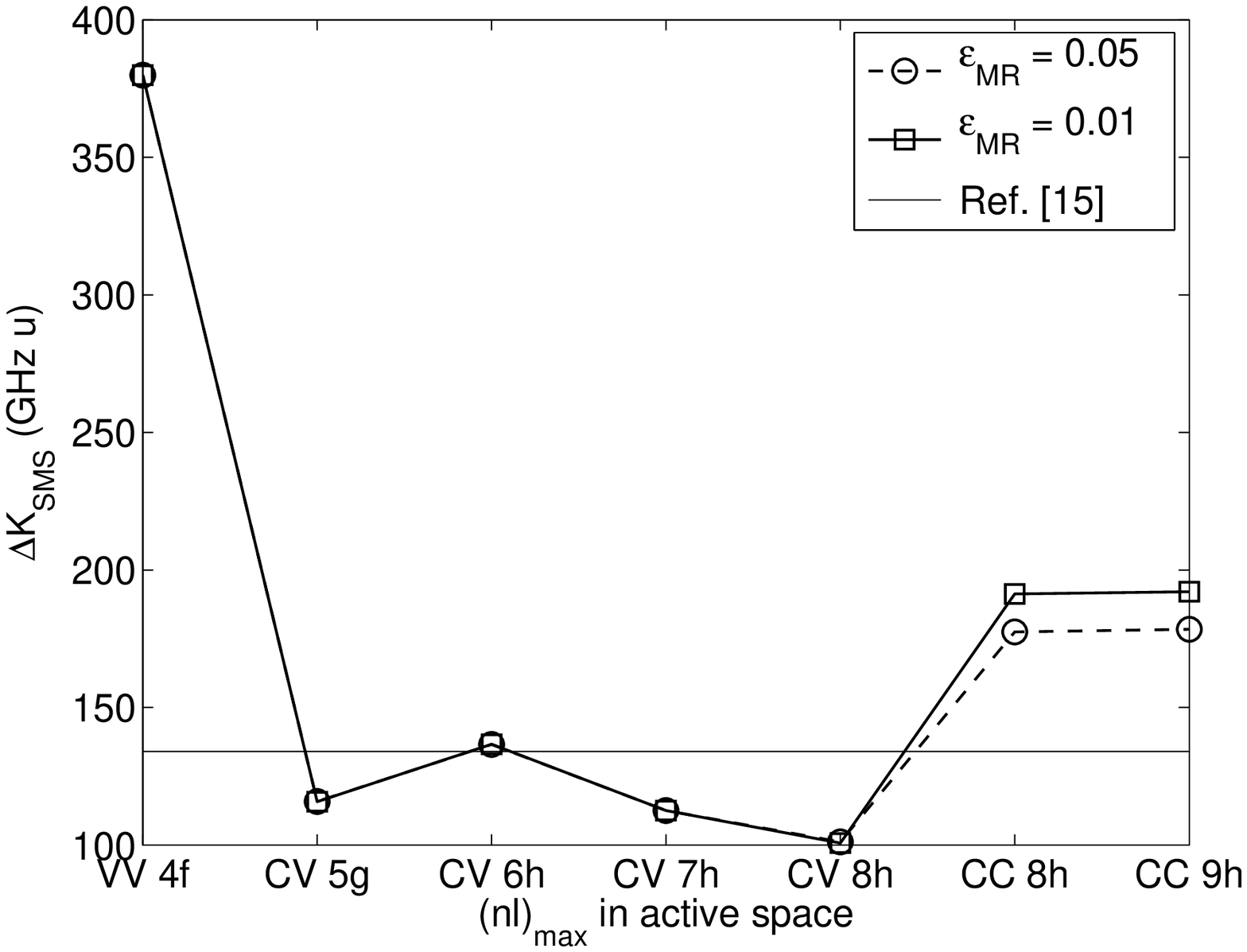}
\end{center}
\end{minipage}
\begin{minipage}[c]{.49\linewidth}
\begin{center}
\includegraphics[scale=0.4]{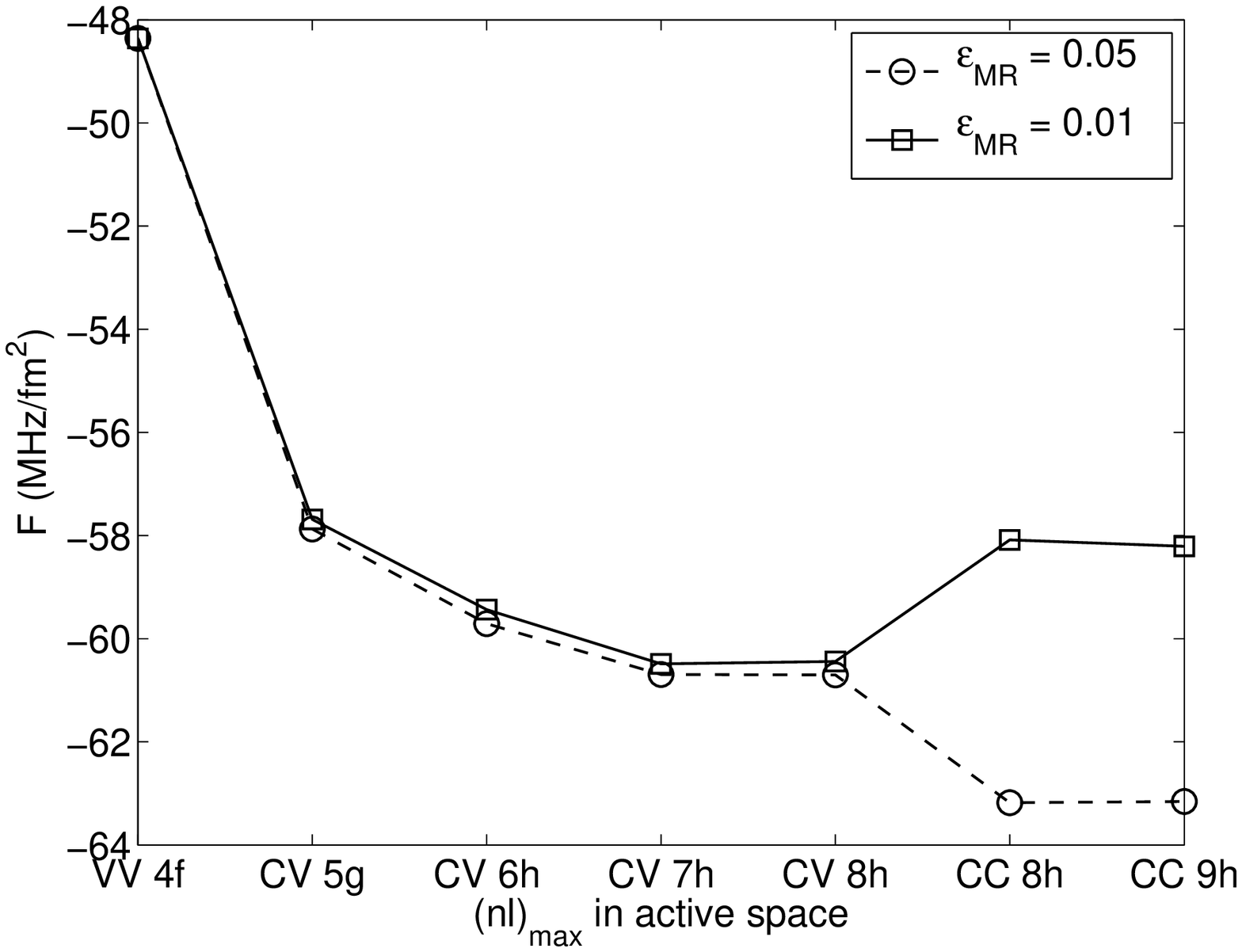}
\end{center}
\end{minipage}
\caption{\small{Transition energy, $\Delta E$ (in cm$^{-1}$), MS factors, $\Delta \tilde{K}_{\text{NMS}}$ and $\Delta \tilde{K}_{\text{SMS}}$ (in GHz~u), and FS factor, $F$ (in MHz/fm$^2$), as functions of the increasing active space for the $3s^2~^1S_0 \rightarrow 3s3p~^1P^o_1$ transition in Mg~I. The correlation models are labelled `VV', `CV' and `CC', and $(nl)_{max}$ denotes the maximal $n$ and $l$ values of the orbitals in the active set. Results are obtained with two MR-cutoff values $\varepsilon_{\text{MR}}$: 0.05 (dashed lines) and 0.01 (solid lines). Comparison of $\Delta E$ with the NIST value~\cite{KRR15}, of $\Delta \tilde{K}_{\text{NMS}}$ with the scaling law from \Eq{eq_sl1}, and of $\Delta \tilde{K}_{\text{SMS}}$ with the benchmark value from \Ref{BFK05}.}}
\label{fig_E_KNMS_KSMS_F}
\end{figure}

Let us now study the impact of the MR-cutoff $\varepsilon_{\text{MR}}$ value, i.e., the size of the MR set, on the accuracy of the transition energy, $\Delta E$ (in cm$^{-1}$), as well as of the MS factors, $\Delta \tilde{K}_{\text{NMS}}$ and $\Delta \tilde{K}_{\text{SMS}}$ (in GHz~u), and the FS factor, $F$ (in MHz/fm$^2$). \figurename{~\ref{fig_E_KNMS_KSMS_F}} displays the convergence plots for the $3s^{2}~^{1}S_{0} \rightarrow 3s3p~^{1}P^{o}_{1}$ transition, as a function of the increasing active space. Two $\varepsilon_{\text{MR}}$ values are considered: 0.05 (dashed lines) and 0.01 (solid lines). For $\varepsilon_{\text{MR}}=0.01$, the MR set of both upper and lower states of this transition are given in \tablename{~\ref{table_MR_composition}}, and the MS and FS results (given in other units) are displayed in \tablename{~\ref{table_KNMS_KSMS_rho}}. For $\varepsilon_{\text{MR}}=0.05$, the reference configurations are [Ne]$\{3s^2,3p^2,3p4p\}$ for the lower state (5 CSFs) and [Ne]$\{3s3p,3s4p,3p3d,3p4s,3d4p\}$ for the upper state (9 CSFs). The size of these MR sets is thus much smaller. The results of $\Delta E$ are compared with the NIST ASD values~\cite{KRR15}, while those of $\Delta \tilde{K}_{\text{NMS}}$ and $\Delta \tilde{K}_{\text{SMS}}$ are respectively compared with the scaling law~\rref{eq_sl1} and with benchmark values from Berengut \textit{et al.}~\cite{BFK05}, in excellent

\clearpage

\onecolumngrid

\twocolumngrid

\noindent agreement with observation (see \tablename{~\ref{table_IS_NMS_SMS_FS}}). These reference values are represented by straight lines in \figurename{~\ref{fig_E_KNMS_KSMS_F}}.

Within the VV model (`VV~$4f$'), the values using $\varepsilon_{\text{MR}}=0.05$ and $0.01$ are the same for each property. Indeed, the computation is performed on the full set of CSFs, before selecting two $\varepsilon_{\text{MR}}$ values leading to different MR sets. Within the VV+CV model (from `CV~$5g$' to `CV~$8h$'), the behaviour of both lines is nearly the~same.

Significant differences occur when CC correlations are added (`CC~$8h$' and `CC~$9h$').
The `CC~$9h$' value of $\Delta E$ is $35\,179$~cm$^{-1}$ with $\varepsilon_{\text{MR}}=0.05$ and $35\,063$~cm$^{-1}$ with $\varepsilon_{\text{MR}}=0.01$, which is closer to the NIST value of $35\,051$~cm$^{-1}$. The same observation holds for $\Delta \tilde{K}_{\text{NMS}}$. The `CC~$9h$' value is $-661$~GHz~u with $\varepsilon_{\text{MR}}=0.05$ and $-566$~GHz~u with $\varepsilon_{\text{MR}}=0.01$, the latter being closer to the scaling law result of $-576$~GHz~u. Equation \rref{eq_sl1}, although only strictly valid in the non-relativistic framework, is used as a reference value since the relativistic effects are expected to be small for $Z=12$. The relativistic corrections to $\Delta \tilde{K}_{\text{NMS}}$ can be deduced with \textsc{Ris}3 by computing the expectation values of the non-relativistic part of the recoil Hamiltonian~\rref{eq_H_MS}, which provides $-576$~GHz~u, reproducing the scaling law result. The relativistic corrections are thus rather small~($2\%$).

The situation is different for $\Delta \tilde{K}_{\text{SMS}}$. The `CC~$9h$' value with $\varepsilon_{\text{MR}}=0.01$ ($192$~GHz~u) is slightly higher than the one with $\varepsilon_{\text{MR}}=0.05$ ($178$~GHz~u). To discriminate between the two results, they are compared with the values from Berengut \textit{et al.}~\cite{BFK05}. In \Ref{BFK05}, $\Delta \tilde{K}_{\text{NMS}}$ is evaluated with the scaling law~\rref{eq_sl1}, and $\Delta \tilde{K}_{\text{SMS}}$ is obtained by the finite-field scaling method. In this technique, the rescaled non-relativistic SMS operator is added to the relativistic many-particle Hamiltonian
\beq
\mathcal{H}_\lambda = \mathcal{H}_0 + \lambda \mathcal{H}_{\text{SMS}} = \mathcal{H}_0 + \lambda \sum_{i<j} \bs{p}_i \cdot \bs{p}_j.
\eeqn{H_lambda}
The eigenvalue problem for Hamiltonian \rref{H_lambda} is solved for various $\lambda$ using a combination of the CI method and many-body perturbation theory (MBPT). Then the level $K_{\text{SMS}}$ factor is evaluated as
\beq
K_{\text{SMS}} = \lim_{\lambda \rightarrow 0} \frac{dE}{d\lambda}.
\eeqn{K_SMS_lambda}
The value of $\Delta \tilde{K}_{\text{SMS}}$ provided by \Ref{BFK05} is $134$~GHz~u,

\clearpage

\onecolumngrid

\begin{table}[ht!]
\caption{\small{Transition energies, $\Delta E$ (in cm$^{-1}$), MS factors, $\Delta \tilde{K}_{\text{NMS}}$ and $\Delta \tilde{K}_{\text{SMS}}$ (in GHz~u), and FS factors, $F$ (in MHz/fm$^{2}$), of the studied transitions in Mg~I. Comparison of $\Delta E$ with values from the NIST database~\cite{KRR15} and theoretical results~\cite{BFK05}. $\Delta \tilde{K}_{\text{NMS}}$ and $\Delta \tilde{K}_{\text{SMS}}$ are respectively compared with values from the scaling law~\rref{eq_sl1} (`Scal.') and with values from \Ref{BFK05}.}}
\begin{center}
\begin{tabular}{l c c c c c c c c c c c c c r r c c c r r}
\hline
\hline
                                                  & & & \mc{4}{c}{$\Delta E$ (cm$^{-1}$)}                       & & & \mc{3}{c}{$\Delta \tilde{K}_{\text{NMS}}$ (GHz~u)} & & & \mc{3}{c}{$\Delta \tilde{K}_{\text{SMS}}$ (GHz~u)} & & & \mc{2}{c}{$F$ (MHz/fm$^{2}$)} \\
\cline{4-7} \cline{10-12} \cline{15-17} \cline{20-21}
Transition                                        & & & CV        & CC        & NIST~\cite{KRR15} & \Ref{BFK05} & & & CV     & CC     & Scal.~\rref{eq_sl1}              & & & \mc{1}{c}{CV} & \mc{1}{c}{CC} & \Ref{BFK05}        & & & CV     & CC                   \\
\hline
$3s^{2}~^{1}S_{0} \rightarrow 3s3p~^{3}P^{o}_{1}$ & & & $21\,970$ & $21\,780$ & $21\,870$         & $21\,794$   & & & $-277$ & $-354$ & $-360$                           & & & $-544$        & $-417$        & $-491$             & & & $-77$  & $-73$                \\
$3s^{2}~^{1}S_{0} \rightarrow 3s3p~^{1}P^{o}_{1}$ & & & $35\,292$ & $35\,063$ & $35\,051$         & $35\,050$   & & & $-492$ & $-567$ & $-576$                           & & & $101$         & $192$         & $\;\;\:134$        & & & $-60$  & $-58$                \\
$3s3p~^{3}P^{o}_{1} \rightarrow 3s4s~^{3}S_{1}$   & & & $19\,474$ & $19\,311$ & $19\,327$         & $19\,332$   & & & $-325$ & $-315$ & $-318$                           & & & $453$         & $416$         & $\;\;\:442$        & & & $40$   & $39$                 \\
$3s3p~^{3}P^{o}_{1} \rightarrow 3p^2~^{3}P_{0}$   & & & $36\,084$ & $35\,857$ & $35\,943$         & $35\,912$   & & & $-501$ & $-570$ & $-591$                           & & & $-17$         & $97$          & $\quad\:27$        & & & $-100$ & $-95$                \\
$3s3p~^{3}P^{o}_{1} \rightarrow 3s3d~^{3}D_{1}$   & & & $26\,324$ & $26\,069$ & $26\,087$         & $26\,085$   & & & $-421$ & $-443$ & $-429$                           & & & $408$         & $403$         & $\;\;\:414$        & & & $26$   & $24$                 \\
$3s3p~^{3}P^{o}_{1} \rightarrow 3s4d~^{3}D_{1}$   & & & $32\,535$ & $32\,173$ & $32\,322$         & $32\,317$   & & & $-504$ & $-505$ & $-532$                           & & & $402$         & $415$         & $\;\;\:403$        & & & $29$   & $27$                 \\
$3s3p~^{1}P^{o}_{1} \rightarrow 3s4d~^{1}D_{2}$   & & & $18\,245$ & $17\,882$ & $18\,084$         & $17\,987$   & & & $-251$ & $-269$ & $-297$                           & & & $-412$        & $-319$        & $-373$             & & & $-5$   & $-7$                 \\
\hline
\hline
\end{tabular}
\end{center}
\label{table_E_KNMS_KSMS_F}
\end{table}

\twocolumngrid

\noindent closer to the result obtained with the higher MR-cutoff. This illustrates again the challenge of providing reliable values of $\Delta \tilde{K}_{\text{SMS}}$. For the SMS factor, the relativistic corrections are small~($2\%$), as expected.

For the FS factor, $F$, the addition of CC correlations leads to two different values at the `CC~$9h$' stage: $-58$~MHz/fm$^2$ for $\varepsilon_{\text{MR}}=0.01$, against $-63$ MHz/fm$^2$ for $\varepsilon_{\text{MR}}=0.05$. Their relative difference can be used to provide an upper bound of the uncertainty on the $F$ factor, equal to $8\%$. This value can be further used in a King plot technique, as the uncertainty on the slope of the straight line, for instance.

The same computation has been performed with an extended MR set. It led to the conclusion that lowering the value of $\varepsilon_{\text{MR}}$ beyond 0.01 does not improve the accuracy of the results. The obtained values with $\varepsilon_{\text{MR}}=0.01$ are thus stable with respect to supplementary correlation effects in the computational procedure. This property holds for all the other transitions studied in this work.

A common observation of the plots displayed in \figurename{~\ref{fig_E_KNMS_KSMS_F}} shows that, although the convergence of the properties is reached within the VV+CV model, the obtained values at that stage are not in excellent agreement with experimental data. This emphasizes the need to include CC excitations in the computational procedure in order to provide more accurate results. This observation is also general for all other transitions studied in this work.

\tablename{~\ref{table_E_KNMS_KSMS_F}} displays the transition energies, $\Delta E$ (in cm$^{-1}$), MS factors, $\Delta \tilde{K}_{\text{NMS}}$ and $\Delta \tilde{K}_{\text{SMS}}$ (in GHz~u), and FS factors, $F$ (in MHz/fm$^{2}$), of the studied transitions in Mg~I. As mentioned in Sec.~\ref{sec:AS}, the labels `CV' and `CC' respectively correspond to the computational procedure in steps 1-2 or steps 1-5.

The values of $\Delta E$ are compared with NIST data~\cite{KRR15} and benchmark results from Berengut \textit{et al.}~\cite{BFK05}. The correction brought by the inclusion of CC correlations is clear. A the `CV' stage all transition energies are over-estimated. In contrast, at the `CC' stage they decrease and become very close to the NIST values. The relative error lies between $0.03\%$ for the $3s^{2}~^{1}S_{0} \rightarrow 3s3p~^{1}P^{o}_{1}$ and $1.12\%$ for the $3s3p~^{1}P^{o}_{1} \rightarrow 3s4d~^{1}D_{2}$ transition, while the calculation performed in \Ref{BFK05} provides relative errors within $0.4\%$ for all considered transitions. The same observation holds for $\Delta \tilde{K}_{\text{NMS}}$. The values are over-estimated at the `CV' stage, and become very close to the scaling law results~\rref{eq_sl2} at the `CC' stage.

Similarly to the study of the $3s^{2}~^{1}S_{0} \rightarrow 3s3p~^{1}P^{o}_{1}$ transition in \figurename{~\ref{fig_E_KNMS_KSMS_F}}, the results of $\Delta \tilde{K}_{\text{SMS}}$ for the other transitions at the `CC' stage are not in better agreement with \Ref{BFK05} than the one obtained at the `CV' stage. They are even less accurate for all considered transitions. These differences represent the major source of discrepancies between this work and experimental values of total IS in Mg~I, as highlighted in \tablename{~\ref{table_IS_NMS_SMS_FS}}.

By contrast, the value of the $F$ factor is not significantly affected by the addition of CC correlations. It varies by a few MHz/fm$^2$ from the `CV' to the `CC' stage.

\tablename{~\ref{table_IS_NMS_SMS_FS}} displays the values of the total IS, NMS, SMS and FS (in MHz) between $^{26}$Mg and $^{24}$Mg of the studied transitions in Mg~I. The NMS and SMS contributions are obtained by multiplying $\Delta \tilde{K}_{\text{NMS}}$ and $\Delta \tilde{K}_{\text{SMS}}$ by the factor $(1/M_{26} - 1/M_{24})$, using Eqs.~\rref{eq_M_26} and \rref{eq_M_24}. The FS are obtained by multiplying $F$ by $\delta \langle r^2 \rangle^{26,24}$, using \Eq{eq_delta_r2_26_24}. The total IS are given by \Eq{eq_Delta_nu_26_24}.

The same conclusions hold for the NMS, SMS and FS values, since they are obtained by multiplying the corresponding electronic factors displayed in \tablename{~\ref{table_E_KNMS_KSMS_F}} by nuclear constants. The NMS results are compared with the scaling law values from \Eq{eq_sl2}, while the SMS results are compared with values extracted from experiments~\cite{SSM93,SHE06,HH78,NGT92,Ha79} and theoretical results of \Ref{BFK05}. The FS contribution is ignored in \Ref{BFK05} for simplicity, since the authors found it to be approximately $20-30$ MHz. Indeed, the FS value is less than the experimental uncertainty in most transitions and is of the order of the error in their SMS calculations. The present results agree with the order of magnitude, but the range of values for the FS is found to be $-6$~-~$+14$ MHz instead.

When considering the total IS, it is worth to observe that the `CV' values are in better agreement with observation than the `CC' ones, for all the studied transitions. Indeed, the errors made on both NMS and SMS within

\clearpage

\onecolumngrid

\begin{table}[ht!]
\caption{\small{Total IS, NMS, SMS and FS (in MHz), between $^{26}$Mg and $^{24}$Mg of the studied transitions in Mg~I. Comparison of NMS with values from the scaling law~\rref{eq_sl2} (`Scal.'). Comparison of IS and SMS with values extracted from experiments ($^a$\Ref{SSM93}, $^b$\Ref{SHE06}, $^c$\Ref{HH78}, $^d$\Ref{NGT92}, $^e$\Ref{Ha79}) and theoretical results~\cite{BFK05}, where the FS contribution is ignored ($20-30$~MHz).}}
\begin{center}
\resizebox{\textwidth}{!}{
\begin{tabular}{l c c r r r c c c r r c c c r r r c c c r r}
\hline
\hline
                                                & & & \mc{4}{c}{IS (MHz)}                                            & & & \mc{3}{c}{NMS (MHz)}                                & & & \mc{4}{c}{SMS (MHz)}                                            & & & \mc{2}{c}{FS (MHz)} \\
\cline{4-7} \cline{10-12} \cline{15-18} \cline{21-22}
Transition                                      & & & \mc{1}{c}{CV} & \mc{1}{c}{CC} & \mc{1}{c}{Expt.} & \Ref{BFK05} & & & \mc{1}{c}{CV} & \mc{1}{c}{CC} & Scal.~\rref{eq_sl2} & & & \mc{1}{c}{CV} & \mc{1}{c}{CC} & \mc{1}{c}{Expt.} & \Ref{BFK05}  & & & CV      & CC        \\
\hline
$3s^2\,^1S_0 \rightarrow 3s3p\,^3P^o_1$         & & & $2643$        & $2482$        & $2683(0)^a$      & $\;2726$    & & & $888$         & $1135$        & $1153$              & & & $1744$        & $1337$        & $1530^a$         & $\;\;\:1573$ & & & $11$    & $10$      \\
$3s^2\,^1S_0 \rightarrow 3s3p\,^1P^o_1$         & & & $1262$        & $1210$        & $1414(8)^b$      & $\;1420$    & & & $1577$        & $1814$        & $1848$              & & & $-324$        & $-612$        & $-434^b$         & $\;\:-428$   & & & $9$     & $8$       \\
$3s3p~^{3}P^{o}_{1} \rightarrow 3s4s~^{3}S_{1}$ & & & $-418$        & $-330$        & $-390(5)^c$      & $-397$      & & & $1041$        & $1009$        & $1019$              & & & $-1453$       & $-1333$       & $-1409^c$        & $-1416$      & & & $-6$    & $-6$      \\
$3s3p~^{3}P^{o}_{1} \rightarrow 3p^2~^{3}P_{0}$ & & & $1674$        & $1529$        & $1810(80)^d$     & $\;1809$    & & & $1606$        & $1827$        & $1895$              & & & $54$          & $-311$        & $-85^d$          & $\quad-86$   & & & $14$    & $13$      \\
$3s3p~^{3}P^{o}_{1} \rightarrow 3s3d~^{3}D_{1}$ & & & $37$          & $125$         & $61(3)^e$        & $\quad\:49$ & & & $1349$        & $1420$        & $1375$              & & & $-1308$       & $-1292$       & $-1314^e$        & $-1326$      & & & $-4$    & $-3$      \\
$3s3p~^{3}P^{o}_{1} \rightarrow 3s4d~^{3}D_{1}$ & & & $324$         & $287$         & $420(20)^d$      & $\;\;\:413$ & & & $1616$        & $1620$        & $1704$              & & & $-1288$       & $-1329$       & $-1284^d$        & $-1291$      & & & $-4$    & $-4$      \\
$3s3p~^{1}P^{o}_{1} \rightarrow 3s4d~^{1}D_{2}$ & & & $2124$        & $1883$        & $2107(15)^c$     & $\;2148$    & & & $804$         & $862$         & $\;\:953$           & & & $1321$        & $1022$        & $1154^c$         & $\;\;\:1195$ & & & $1$     & $1$       \\
\hline
\hline
\end{tabular}
}
\end{center}
\label{table_IS_NMS_SMS_FS}
\end{table}

\twocolumngrid

\noindent the VV+CV model seem to `accidentally' cancel, providing more accurate values for the total IS. By contrast, within the VV+CV+CC model the NMS values are closer to the scaling law results, but the SMS values are not improved in comparison. Summing up NMS and SMS leads thus to less accurate results for the total IS.

Compared to the values from \Ref{BFK05}, the total IS is in less good agreement with observation for all studied transitions, whether CC correlations are included or not. Indeed, the MBPT+CI method is known to be the most accurate computational technique for one- and two-valence-electron atoms. Nevertheless, these results show that CC effects need to be accounted for in the computational strategy, in order to improve the values of $\Delta E$ and $\Delta \tilde{K}_{\text{NMS}}$ for each of the studied transitions in Mg~I.

%%%%%%%%%%%%%%%%%%%%%%%%%%%%%%%%%%%%%%%%%%%%%%%%%%%%%%%%%%%%%%%%%%%%%%%%%%%%%%%%%%%%%%%%%%%%%%%%%%%%%%%%%%%%%%%%%%%%%%%%%%
%%%%%%%%%%%%%%%%%%%%%%%%%%%%%%%%%%%%%%%%%%%%%%%%%%%%%%%%%%%%%%%%%%%%%%%%%%%%%%%%%%%%%%%%%%%%%%%%%%%%%%%%%%%%%%%%%%%%%%%%%%

\section{Conclusion}
\label{sec:conc}

The present work describes an \textit{ab initio} method for the relativistic calculation of the IS in many-electron atoms using the MCDHF approach. The accuracy of the computational procedure is tested by estimating the energy shifts of the $^{26}$Mg$-^{24}$Mg pair of isotopes, for several well-known transitions in Mg~I.

Different models for electron correlation are adopted. Within each model, the convergence of the level MS factors and the electronic probability density at the origin, as a function of the increasing active space, is studied for the $3s^{2}~^{1}S_{0} \rightarrow 3s3p~^{1}P^{o}_{1}$ transition. It is shown that small variations in the level values due to correlation effects can lead to a significant variation in the differential values, highlighting the challenge in providing accurate results for the SMS factors with this computational approach. The impact of the MR-cutoff value on the accuracy of the transition energy and the MS and FS electronic factors is investigated as a function of the increasing active space, for the same transition. It leads to the conclusion that extending the MR set beyond a certain MR-cutoff value does not improve the accuracy of the results.

The study of the electronic factors for other transitions in Mg~I shows that CC correlation needs to be accounted for in the computational strategy, in order to obtain accurate values for the transition energies and the NMS factors. The convergence of the results when including an additional orbital layer optimised on CC substitutions from the Mg$^{2+}$ core is highly satisfactory. By contrast, in comparison with benchmark calculations from Berengut \textit{et al.}~\cite{BFK05}, the accuracy of the SMS factor values is not improved when CC contributions are added.

Total IS, NMS, SMS and FS are computed between $^{26}$Mg and $^{24}$Mg for the studied transitions in Mg~I. The agreement of the numerical results is found to be good for all transitions. It is surprisingly better for the VV+CV model, although the transition energies and the NMS factors are less accurate than in the VV+CV+CC model. In the former, the errors made on NMS and SMS, cancel each other out `accidentally', providing more accurate values for the total IS. Nevertheless, for both correlation models, the present accuracy is in particular high enough for the purposes of resolving systematic errors in the search for the fine-structure constant variation, and for studies of the isotopic evolution of the universe~\cite{BFK05}.

A possible way to improve the accuracy of the results is the use of the partitioned correlation function interaction (PCFI) approach~\cite{VRJ13}. It is based on the idea of relaxing the orthonormality restriction on the orbital basis, and breaking down the very large calculations in the traditional multiconfiguration methods into a series of smaller parallel calculations. This method is very flexible for targeting different electron correlation effects. CC effects in IS factors could be then treated more accurately and efficiently with the use of this technique. Additionally, electron correlation effects beyond the SD-MR model (such as triple and quadruple excitations) can be included perturbatively. Work is being done in these directions.

\vspace{-0.5cm}

%%%%%%%%%%%%%%%%%%%%%%%%%%%%%%%%%%%%%%%%%%%%%%%%%%%%%%%%%%%%%%%%%%%%%%%%%%%%%%%%%%%%%%%%%%%%%%%%%%%%%%%%%%%%%%%%%%%%%%%%%%
%%%%%%%%%%%%%%%%%%%%%%%%%%%%%%%%%%%%%%%%%%%%%%%%%%%%%%%%%%%%%%%%%%%%%%%%%%%%%%%%%%%%%%%%%%%%%%%%%%%%%%%%%%%%%%%%%%%%%%%%%%

\begin{acknowledgments}
This work has been partially supported by the Belgian F.R.S.-FNRS Fonds de la Recherche Scientifique (CDR J.0047.16) and the BriX IAP Research Program No. P7/12 (Belgium). L.F. acknowledges the support from the FRIA. J.E. and P.J. acknowledge financial support from the Swedish Research Council (VR), under contract 2015-04842.
\end{acknowledgments}

%%%%%%%%%%%%%%%%%%%%%%%%%%%%%%%%%%%%%%%%%%%%%%%%%%%%%%%%%%%%%%%%%%%%%%%%%%%%%%%%%%%%%%%%%%%%%%%%%%%%%%%%%%%%%%%%%%%%%%%%%%
%%%%%%%%%%%%%%%%%%%%%%%%%%%%%%%%%%%%%%%%%%%%%%%%%%%%%%%%%%%%%%%%%%%%%%%%%%%%%%%%%%%%%%%%%%%%%%%%%%%%%%%%%%%%%%%%%%%%%%%%%%

\end{document}